\documentclass[letterpaper,12pt]{JHEP3}
\usepackage{epsfig}
\usepackage{amssymb}
\def\be{\begin{equation}}
\def\ee{\end{equation}}
\def\bea{\begin{eqnarray}}
\def\eea{\end{eqnarray}}

\title{Geometrothermodynamics of asymptotically anti--de Sitter black holes}
\author{Hernando Quevedo \\
Instituto de Ciencias Nucleares\\
Universidad Nacional Aut\'onoma de M\'exico  \\
E-mail: \email{quevedo@nucleares.unam.mx}}
\author{Alberto S\'anchez \\
Instituto de Ciencias Nucleares\\
Universidad Nacional Aut\'onoma de M\'exico  \\
E-mail: \email{asanchez@nucleares.unam.mx}}
\abstract{
We apply the formalism of geometrothermodynamics to 
the case of black holes with cosmological constant 
in four and higher dimensions.
We use a thermodynamic 
metric which is invariant with respect to Legendre 
transformations and determines the geometry of 
the space of equilibrium states. For all known black holes
in higher dimensions, we show that the 
curvature scalar of the thermodynamic metric in all the
cases is proportional to the heat capacity. As a consequence,
phase transitions, which correspond to divergencies of 
the heat capacity, are represented geometrically as 
true curvature singularities. We interpret this as a further
indication that the curvature of the thermodynamic metric 
is a measure of thermodynamic interaction.}

\preprint{{}}
\begin{document}
%\maketitle
%\begin{keyword}
% Equilibrium thermodynamics \sep contact structure \sep Riemannian structure \sep 
% Weinhold's metric
%\end{keyword}

% main text
\section{Introduction}
\label{sec:int}
% Differential geometry is a very important tool of modern science, 
% specially of mathematical physics and its applications in physics,
% chemistry and engineering. 
One of the most interesting results of modern theoretical physics is its
direct relation with many areas of mathematics like differential geometry.
General relativity, for instance, can be considered mathematically 
as an application of differential geometry. Once a metric is given which is
compatible with a torsion-free 
connection and satisfies Einstein's equations, the
corresponding curvature turns out to be a measure of gravitational interaction.
This is a fascinating result that combines apparently different concepts 
of geometry and physics, allowing us to study gravity by measuring the curvature
of spacetime. In fact, this result can conceptually be generalized to include 
all the field interactions that are known in nature. The electromagnetic, weak,
and strong interactions can classically be described by using 
the Minkowski metric and a gauge connection. In all the cases, the resulting 
gauge curvature can be considered as a measure of the corresponding field interaction
(see, for instance, \cite{frankel}).

During the last few decades several attempts have been made in 
order to introduce differential geometric concepts in ordinary thermodynamics.  
Hermann \cite{her} formulated the concept of thermodynamic phase space
as a differential manifold with a natural contact structure. 
In the thermodynamic phase space there exists a special subspace of 
thermodynamic equilibrium states. Weinhold \cite{wei1} proposed an alternative approach
in which in the space of equilibrium
states  a metric is introduced {\it ad hoc} as the Hessian of the internal energy.
In an attempt to formulate the concept of thermodynamic length, Ruppeiner
\cite{rup79} introduced a metric which is conformally equivalent to Weinhold's metric.
The study of the relation between the phase space and the metric structures 
of the space of equilibrium states led to the result that Weinhold's and Ruppeiner's
thermodynamic 
metrics are not invariant under Legendre transformations \cite{sal83,mru90}, i.e.
the geometric properties of the space of equilibrium states are different 
when different thermodynamic potentials are used. This result clearly contradicts
ordinary equilibrium thermodynamics which is manifestly Legendre invariant.
Moreover, the question whether the curvature of the space of equilibrium 
states can be considered as a measure for thermodynamic interaction remained 
unanswered. This was particularly clear in the case of the thermodynamics of black holes
where a flat thermodynamic metric can be transformed into a non-flat metric by 
means of a Legendre transformation \cite{quev08}.  

Recently, the formalism of geometrothermodynamics (GTD) was developed in order to 
unify in a consistent manner the geometric properties of the phase space and the
space of equilibrium states \cite{quev07}. Legendre invariance plays an important
role in this formalism. In particular, it allows us to derive Legendre invariant
metrics for the space of equilibrium states. It has been shown that there 
exist thermodynamic metrics which correctly describe the thermodynamic behavior
of the ideal gas and the van der Waals gas. In fact, for the ideal gas the 
curvature vanishes whereas for the van der Waals gas the curvature is non-zero
and diverges only at those points where phase transitions take place. 
Moreover, in the case of black hole thermodynamics in four dimensions we have
shown recently \cite{aqs08}
that there exists a thermodynamic metric with non-vanishing curvature which 
correctly describes the thermodynamic properties of those black holes. 
The main goal of the present work is to show that for all known 
asymptotically anti-de Sitter black holes in all dimensions there exists
a thermodynamic metric with non-zero curvature which correctly describes
the structure of phase transitions as dictated by the corresponding
heat capacity. Consequently, our main result is that the curvature of 
the space of equilibrium states can be used in a general manner 
as a measure of the thermodynamic interaction of black holes. 

The study of classical gravitational configurations on a background with 
cosmological constant has been intensified in the last few years. First, 
cosmological observations indicate that a positive cosmological constant could be 
responsible for the present acceleration of the Universe.  
On the other hand, a negative cosmological constant plays a distinguished role 
in the conjectured AdS/CFT correspondence, according to which 5-dimensional 
solutions 
of Einstein equations with negative cosmological constant can be used to derive
certain statements about quantum field theory in four dimensions.
In this context, the thermodynamic properties of black holes in an AdS background
acquire especial importance since they give information about quantum field theory 
at non-zero temperature. Charged, rotating black holes in an AdS background are known
explicitly only in four \cite{carter} and five dimensions \cite{haw99}. Moreover, 
Reissner-Nordstr\"om-AdS and Kerr-AdS black holes are known in all dimensions 
\cite{rnall, kerrall}. The thermodynamics of these higher dimensional black holes 
has been a subject of intensive investigation due to its importance in the context 
of the AdS/CFT conjecture \cite{cal00,gib05,ash07}. In 2+1 gravity, 
the BTZ black hole presents an interesting phase transitions structure 
\cite{cai99}. 
Charged topological AdS black holes
and their phase transitions were analyzed in \cite{jose00}. 
One of the interesting results is
that all the intrinsic parameters that characterize black holes in higher dimensions 
can be treated, with certain care, as thermodynamic variables of ordinary thermodynamics.

This paper is organized as follows.  In Section \ref{sec:gtd} we review 
the general formalism of GTD for black holes in arbitrary dimensions 
and introduce a Legendre 
invariant metric in the thermodynamic phase space which is used to
generate the geometric structure of the space of equilibrium states. 
In Sections \ref{sec:highrn} and \ref{sec:highkerr} we investigate the 
structure of the phase transitions of the Reissner-Nordstr\"om AdS 
and Kerr AdS black holes, respectively, and show the points where 
phase transitions occur are characterized by curvature singularities 
of the thermodynamic metric.  
In the final Section \ref{sec:con} we discuss
our results. 
Throughout this paper we use units in which $G=c=k_{_B}=\hbar =1$.

%%%%%%%%%%%%%%%%%%%%%%%%%%%%%%%%%%%%%%%%%%%%%%%%%%%%%%
%%%%%%%%%%%%%%%%%%%%%%%%%%%%%%%%%%%%%%%%%%%%%%%%%%%%%%
\section{Geometrothermodynamics}
\label{sec:gtd}

In order to describe a thermodynamic system with $n$ degrees of freedom,
we consider in GTD the thermodynamic phase space which is defined mathematically as 
a Riemannian contact manifold $({\cal T}, \Theta, G)$, where 
${\cal T}$ is a $(2n+1)-$dimensional manifold, $\Theta$ is a linear differential 
1-form satisfying the condition $\Theta \wedge (d\Theta)^n \neq 0$, 
and $G$ is a non-degenerate, Legendre invariant metric on ${\cal T}$.
Here $\wedge$ represents the exterior product, $d$ is the exterior derivative,
and 
$(d\Theta)^n = d\Theta \wedge ... \wedge d\Theta$ ($n-$times). 
The submanifold ${\cal E}\subset {\cal T}$ defined by means of 
a smooth embedding mapping  
$ \varphi : \   {\mathcal E} \  \longrightarrow {\mathcal T}$ such that 
the pullback $\varphi^*(\Theta)=0$ is called the space of thermodynamic 
equilibrium states. A Riemannian structure $g$ is induced naturally in ${\cal E}$
by means of $g=\varphi^*(G)$. It is then expected in GTD \cite{quev07} that the physical 
properties of a thermodynamic system in a state of equilibrium can be described 
in terms of the geometric properties of the corresponding space of equilibrium 
states ${\cal E}$.

To be more specific we introduce in the phase space ${\cal T}$ the coordinates
$Z^A=(\Phi, E^a, I^a)$ with $A=0,...,2n$, and $a=1,...,n$. In ordinary 
thermodynamics, $\Phi$ corresponds to the thermodynamic potential, and $E^a$ and
$I^a$ are the extensive and intensive variables, respectively. The fundamental differential
form $\Theta$ can then be written in a canonical manner as 
$\Theta = d\Phi - \delta_{ab} I^a d E^b$, where $\delta_{ab}$ is the Euclidean metric.
The metric components in ${\cal T}$ can be in general arbitrary $C^2-$functions of the coordinates, 
i. e., $G_{AB}=G_{AB}(Z^C)$. This arbitrariness is restricted by the 
condition that $G$ must be invariant with respect to Legendre transformations. 
This  is a necessary condition for our description of thermodynamic systems 
to be independent of the thermodynamic potential. This implies that ${\cal T}$ must 
be a curved manifold \cite{quev07} 
because the special case of a metric with vanishing curvature turns out to be
non Legendre invariant. Although in general any $n-$dimensional subset of the set of coordinates $Z^A$ can 
be used to coordinatize the submanifold ${\cal E}$, for the sake of simplicity we choose the subset 
$E^a$ as coordinates of ${\cal E}$. Then the smooth mapping $\varphi: {\cal E} \longrightarrow {\mathcal T}$ 
is given in terms of coordinates as $ \varphi :  \{E^a\} \longmapsto Z^A=\{\Phi(E^a), E^a, I^a(E^a)\}$.
Consequently, the condition $\varphi^*(\Theta)=0$ can be written as the expressions
\be
d\Phi = \delta_{ab} I^a d E^b \ , \qquad \frac{\partial\Phi}{\partial E^a} = 
\delta_{ab} I^b \ ,
\ee
which in ordinary thermodynamics correspond to the first law of thermodynamics and the conditions 
for thermodynamic equilibrium, respectively. We see that the specification of the mapping $\varphi$ 
includes the specification of the relationship $\Phi=\Phi(E^a)$ that is nothing more but the fundamental 
equation from which all the information about a thermodynamic system can be obtained \cite{callen}.
The second law of thermodynamics is implemented in GTD as the convexity condition
\be
\frac{\partial^2\Phi}{\partial E^a \partial E^b} \geq 0 \ . 
\ee

To complete our construction we need a metric $G$. There is a large arbitrariness in the selection of this 
metric since at this level it is only demanded that it satisfies the condition of Legendre invariance. For the
sake of simplicity we will use the following choice
\be
G= (d\Phi - \delta_{ab} I^a d E^b)^2 +  (\delta_{ab}E^a I^b)(\eta_{cd} dE^c dI^d)
\ ,\qquad \eta_{ab}={\rm diag}(-1,1,...,1) \ ,
\label{giiup}
\ee
where $\eta_{ab}$ is a pseudo-Euclidean metric in ${\cal E}$. This metric is a slight modification 
of the metric $G^{II}$ presented in \cite{quev07} which was used there to generate 
the simplest Legendre invariant generalizations of Weinhold's and Ruppeiner's thermodynamic metrics.
 It is easy to show that the metric (\ref{giiup}) is 
invariant with respect to the Legendre transformation  \cite{arnold}
\be
\{\Phi, E^a,I^a\}\longrightarrow \{\tilde \Phi, \tilde E ^a, \tilde I ^ a\}
\ee
\be
 \Phi = \tilde \Phi - \delta_{ab} \tilde E ^a \tilde I ^b \ ,\quad
 E^a = - \tilde I ^ {a}, \ \  
 I^{a} = \tilde E ^ a \ ,
 \label{leg}
\ee
and when ``projected" on ${\cal E}$ by means of $g=\varphi^*(G)$ generates the thermodynamic metric
\be
g=\left(E^c\frac{\partial \Phi}{\partial E^c}\right)
\left(\eta_{ab}\delta^{bc}\frac{\partial^2\Phi}
{\partial E^c \partial E^d}
dE^a dE^d\right) \ .
\label{giidown}
\ee 
Once the fundamental equation $\Phi=\Phi(E^a)$ is known for a given thermodynamic system, the explicit form
of the thermodynamic metric $g$
 can easily be computed. If the curvature of the thermodynamic metric is to be
considered as a measure of the thermodynamic interaction, the metric (\ref{giidown}) should be flat only for
 systems with no thermodynamic interaction. Moreover, phase transitions associated with divergencies of the
 thermodynamic interaction 
 should correspond to curvature singularities. We will see that the metric (\ref{giidown}) satisfies these
conditions in the case of thermodynamic systems represented by black holes.

%%%%%%%%%%%%%%%%%%%%%%%%%%%%%%%%%%%%%%%%%%%%%%%%%%%%%
%%%%%%%%%%%%%%%%%%%%%%%%%%%%%%%%%%%%%%%%%%%%%%%%%%%%%
\section{Four dimensional Kerr-Newman-AdS black hole}
\label{sec:knads4}
In the Einstein-Maxwell theory with cosmological constant $\Lambda$, which follows from the 
action 
\be
S_{_{EM}}= \frac{1}{16\pi} \int_{M^4} d^4 x [-{\rm det}(g_{\mu\nu})]^{1/2}\left( R - F_{\mu\nu}F^{\mu\nu} - 2\Lambda \right) \ ,
\ee
the most general solution representing a black hole configuration is given by the Kerr-Newman-AdS solution \cite{carter}
that in Boyer-Lindquist-like coordinates can be expressed as 
\be
ds^2 = -\frac{\Delta_r}{\rho^2}\left(dt-\frac{a\sin^2\theta}{\Xi} d\varphi\right)^2 
+\frac{\Delta_\theta\sin^2\theta}{\rho^2} \left(a dt - \frac{r^2+a^2}{\Xi} d\varphi\right)^2
+{\rho^2}\left(\frac{dr^2}{\Delta_r} +\frac{d\theta^2} {\Delta_\theta} \right)
\ee
where
\be
\Delta_r =(r^2+a^2)\left(1+\frac{r^2}{l^2}\right) - 2mr + q^2  \ ,\qquad 
\Delta_\theta = 1 -\frac{a^2}{l^2}\cos^2\theta \ ,
\ee
\be
\rho^2 = r^2 + a^2\cos^2\theta \ , \qquad \Xi = 1 - \frac{a^2}{l^2}\ .
\ee
This solution describes the gravitational field of a charged, rotating 
 black hole with cosmological constant $\Lambda = -3/l^2$, where $l$ is 
the curvature radius of the AdS spacetime. The electromagnetic potential $A_\mu$ 
is given as 
\be
A_t= -\frac{qr}{\rho^2}\ ,\qquad A_\varphi = \frac{aqr\sin^2\theta}{\rho^2\Xi}
\ee
with the angular, magnetic component $A_\varphi$ that appears as a consequence of 
the rotation of the black hole. The horizons are determined by the (positive)
roots of the equation $\Delta_r=0$. In particular, the outer horizon is situated
at $r=r_+$ and corresponds to the largest root.  

The physical properties of this spacetime can 
be understood by considering the physical parameters entering the metric functions. 
The area of the horizon is a well-defined geometric parameter 
$ A=\int \sqrt{g_{\varphi\varphi} g_{\theta\theta}}d\varphi d\theta$  that can easily be
calculated at $r=r_+$ and yields $A= 4\pi (r_+^2+a^2)/\Xi$.   
The surface gravity $\kappa$ can be derived, modulo a trivial additive constant, 
from the equation $k^a k^b_{;a} -\kappa k^b=0$ evaluated at the horizon for 
a timelike Killing vector field $k^a$. In the case under consideration we
have that $k^a\partial_a =\partial_t + \Omega \partial_\varphi$, where $\Omega$ is
the angular velocity measured by a non-rotating observer at infinity,
so that the surface gravity is given as 
$\kappa = [3r_+^4+(a^2+l^2)r_+^2-(a^2+q^2)l^2]/[2l^2r_+(r_+^2+a^2)]$. The
situation is more complicated in the case of the physical mass (total energy) $M$, 
angular momentum $J$, and electric charge $Q$,
 because these parameters are usually defined for asymptotically flat 
spacetimes. For asymptotically AdS spacetimes several definitions are possible 
and the issue has been clarified only recently by using the laws of black holes 
thermodynamics and the formalism of isolated horizons \cite{gib05,ash07}. 
It turns
out that it is necessary to measure the angular velocity with respect to 
an observer which is not rotating at infinity \cite{cal00}. Then the computation 
of the intrinsic physical parameters results in 
\be 
M= \frac{m}{\Xi^2}\ ,\qquad J= \frac{am}{\Xi^2}\ , \qquad
Q = \frac{q}{\Xi} \ .
\label{physpar}
\ee

The connection to thermodynamics arises when one considers the Bekenstein-Hawking
entropy in terms of the horizon area, i.e. $S= A/4$. It is then easy to 
derive the generalized Smarr formula for the KN-AdS black hole 
\be
M^2=J^2 \left(\frac{1}{l^2}+ \frac{\pi}{S}\right)
+ \frac{S^3}{4\pi^3}\left(\frac{1}{l^2}+ \frac{\pi}{S} + \frac{\pi^2Q^2}{S^2}
\right)^2 \ ,
\label{feknads4}
\ee
which is the fundamental thermodynamic equation. It relates the total energy $M$
of the black hole with the extensive variables $S$, $Q$, and $J$. As in ordinary
thermodynamics, in GTD it is the fundamental equation from which all the 
thermodynamic information can be derived. With the choice $E^a =\{S,Q,J\}$, the 
corresponding intensive variables become $I^a=\{T,\phi,\Omega\}$, where
$\phi$ is the electric potential and $\Omega$ is the angular velocity. 
Furthermore, with this choice $M$ corresponds to the thermodynamic
potential. In this way, we have introduced all the coordinates $Z^A= \{M,S,Q,J,T,
\phi, \Omega\}$ of the 7-dimensional thermodynamic phase space ${\cal T}$
which, 
according to Eq.(\ref{giiup}), becomes a Riemannian manifold with metric 
\be
G=(dM-TdS  - \phi d Q - \Omega dJ )^2 +
(ST+ \phi  Q+ \Omega J )\left(-dSdT + dQ d \phi + dJ d\Omega \right) \ . 
\label{giiknads4}
\ee
This is a non-degenerate metric with 
${\rm det}(G_{AB})= (ST+\Omega J + \phi  Q)^4/16$ 
and non-zero curvature. Moreover, the contact structure of ${\cal T}$ is generated
by the fundamental form $\Theta =dM-TdS  - \phi d Q - \Omega dJ$. 
At the level of the phase space ${\cal T}$, the metric (\ref{giiknads4})
plays an auxiliary role in the sense that it generates a Legendre invariant metric 
for the space of equilibrium states ${\cal E}$ with coordinates $\{E^a\}$. 
To this end, we introduce the smooth mapping 
\be
\varphi: \{S,Q,J\} 
\longmapsto \left\{ M(S,Q,J), S,Q,J, T(S,Q,J), \phi(S,Q,J),\Omega(S,Q,J)
\right\}
\ee
by using the fundamental equation (\ref{feknads4}) and the condition  
$\varphi^*(\Theta)=0$ so that on ${\cal E}$ it corresponds to the 
first law of black hole thermodynamics $dM=TdS  +\phi d Q + \Omega dJ$.
This, in turn, can be used to compute the dual intensive variables 
corresponding to the temperature
\be
T=\frac{\partial M}{\partial S} = 
\frac{S^2}{8M\pi^3}\left(\,\frac{1}{l^2}+\frac{\pi}{S}+\frac{\pi^2 Q^2}{S^2}\,\right)\left(\,\frac{3}{l^2}+\frac{\pi}{S}
-\frac{\pi^2 Q^2}{S^2}\,\right)
-\frac{\pi J^2}{2MS^2}
 \ ,
\ee
the electric potential,
\be
\phi = \frac{\partial M}{\partial Q} = 
\frac{QS}{2M\pi}\left(\,\frac{1}{l^2}+\frac{\pi}{S}+\frac{\pi^2 Q^2}{S^2}\,\right)
 \ ,
\ee
and angular velocity 
\be
\Omega = \frac{\partial M}{\partial J} = 
\frac{J}{M}\left(\frac{1}{l^2}+\frac{\pi}{S}\right)
 \ .
\ee
Moreover, according to Eq.(\ref{giidown}), the metric structure of ${\cal E}$ is
given as
\be
g=(SM_S + QM_Q + JM_J)\left(
\begin{array}{ccc}%[pos]
-M_{SS}& 0 & 0 \\
0 &  M_{QQ} &  M_{QJ} \\
0 &  M_{QJ} & M_{JJ} 	
\end{array}
\right) \ ,
\label{gknads}
\ee
where subscripts represent partial derivative with respect to the corresponding
coordinate. Notice that no cross terms of the form $g_{SQ}$ or $g_{SJ}$ appear in 
this expression, which would be proportional to $M_{SQ}$ or $M_{SJ}$, respectively.
This is due to the special choice of the auxiliary metric 
(\ref{giiknads4}). Indeed, the minus sign in front of the term $dSdT$ in $G$ 
leads to the disappearance of the cross terms of $g=\varphi^*(G)$ that involve 
the coordinate $S$, i. e., $g_{SJ}$ and $g_{SQ}$. Our 
choice of $G$ is in agreement with the condition of Legendre invariance and
was inspired by inspecting the expression of the scalar curvature $R$. 
In fact, $R$ always contains the determinant of the metric $g$ in the 
denominator and, therefore, the zeros of $\det(g)$ could lead to curvature 
singularities (if those zeros are not canceled by the zeros of the numerator). 
On the other hand, as we will show below, the locations of the divergencies 
of the heat capacity  coincide with the zeros of $M_{SS}$. 
Then, the choice of the metric (\ref{giiknads4}) has the purpose of generating 
a metric $g$ whose determinant is proportional to $M_{SS}$, leading 
to a one-to-one correspondence between the divergencies of the heat capacity and 
singularities of the scalar curvature.

In the thermodynamics of black holes, phase transitions must play an important role.
Due to the absence of a realistic, microscopic model for the entropy of black 
holes, a problem which is related to the absence of a theory of quantum gravity, 
there is no unanimity about the definition of phase transitions \cite{rup08}. 
Nevertheless, one can adopt the point of view of ordinary thermodynamics and search
for singular points in the behavior of thermodynamic variables. Such an approach 
was realized by Davies \cite{davies}, showing that divergencies in the heat capacity 
indicate the points where phase transitions occur. We follow Davies' approach in
this work. From the fundamental equation (\ref{feknads4}) it is straightforward
to compute the heat capacity for the KN-AdS black hole:
\bea
\protect\label{heatknads4}
C & = & T \frac{\partial S}{\partial T} = \frac{M_S}{M_{SS}}  \\
 & = & \frac{S 
\left( \frac{1}{l^2}+\frac{\pi}{S}+\frac{\pi^2 Q^2}{S^2}\,\right)
\left(\,\frac{3}{l^2}+\frac{\pi}{S}-\frac{\pi^2 Q^2}{S^2}\,\right)
-\frac{4\pi^4 J^2 }{S^3}}
{  
\left(\frac{1}{l^2}+\frac{\pi}{S}+\frac{\pi^2 Q^2}{S^2} \right)
\left( \frac{6}{l^2}+\frac{\pi}{S}\right)
-\left(\frac{\pi}{S}+\frac{2\pi^2 Q^2}{S^2} \right)
\left( \frac{3}{l^2}+\frac{\pi}{S}- \frac{\pi^2  Q^2}{S^2}\right)
+\frac{8\pi^3}{S^2}\left(\frac{\pi J^2}{S^2}-ST^2\right) }
\nonumber
\eea
Phase transitions are then determined by the roots of the denominator of $C$, i.e.
$M_{SS} =0$.

On the other hand, for the curvature of the metric (\ref{gknads}) to be a measure 
of the thermodynamic interaction in the KN-AdS black hole, it must reproduce 
the phase transitions structure dictated by the heat capacity (\protect\ref{heatknads4}).
To verify this property in an invariant manner we compute the scalar curvature
of the thermodynamic metric  (\ref{gknads}), and notice that its denominator is given by
\be
D_R= 4 (SM_S + QM_Q + JM_J)^3 ( M_{QJ}^2 - M_{QQ}M_{JJ} ) ^3 M_{SS}^2 \ ,
\ee
whereas the numerator is a rather cumbersome expression that can not be written 
in a compact form. We see that the denominator is proportional to the determinant
of the metric (\ref{gknads}). At first sight, 
the singular points of the heat capacity that are 
situated at $M_{SS}=0$ correspond to true curvature singularities where the volume
element vanishes. However, this is valid only if the numerator of the scalar curvature
does not eliminate the zeros of the denominator. A numerical analysis shows that
in fact the singularities of the heat capacity coincide with the singularities of
the scalar curvature. We first fix the value of the cosmological constant and the
entropy in such a way that we get for the charge and angular momentum an interval   
where the heat capacity diverges. The same process is then repeated for different 
combinations of values for the cosmological constant and entropy. As a result we 
find all the intervals where divergences occur. Around the divergent points of 
the heat capacity we then investigate the behavior of the thermodynamic scalar 
curvature and confirm that it becomes singular. We also noticed that the 
singular points coincide with the zeros of $M_{SS}$ so that in fact  
the curvature singularities are situated at the points where phase transitions
take place. The characteristic behavior of the heat capacity and curvature 
is depicted in figures \ref{fig:knlc} and \ref{fig:knlr}.
\DOUBLEFIGURE{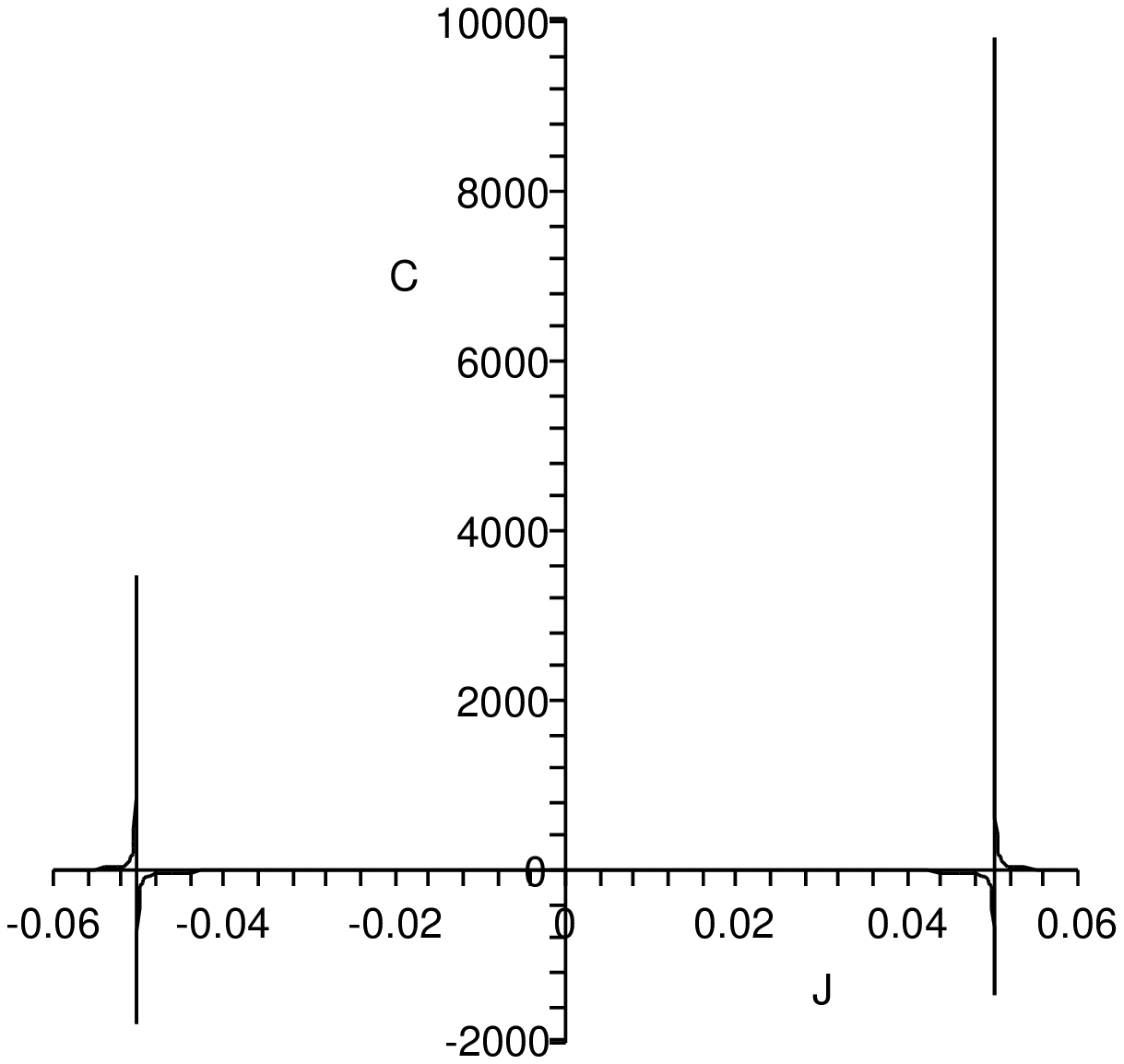,width=7cm}{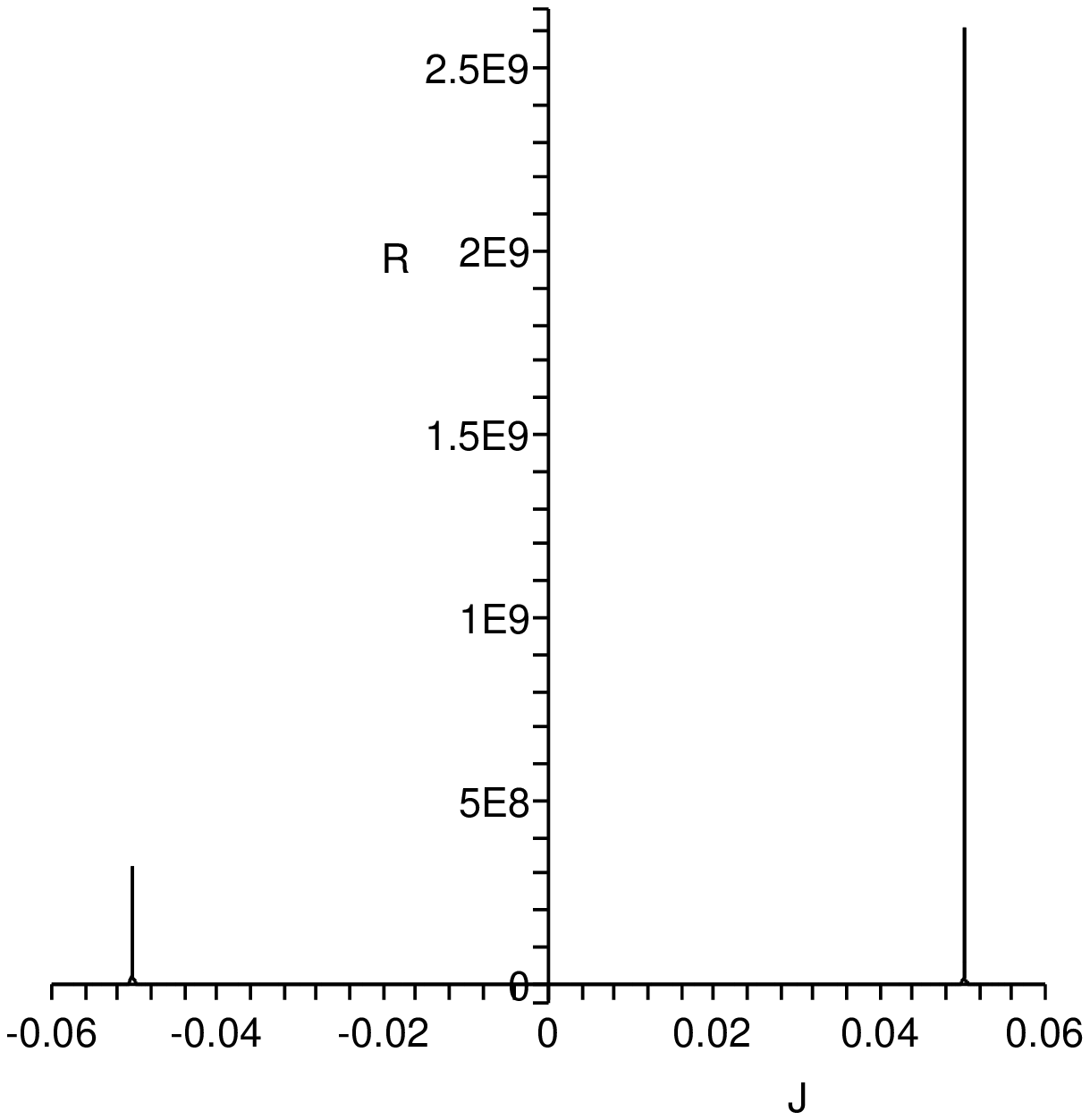,width=7cm}
{Characteristic behavior of the heat capacity $C$ in terms of the angular momentum $J$. 
The chosen values are $\Lambda=-1$, $S=1$, and $Q=0.01$. The divergencies 
indicate points of phase transitions.
\protect\label{fig:knlc}}
{The thermodynamic scalar curvature $R$ in terms of the angular momentum $J$. 
The values of the remaining parameters are as in figure \protect\ref{fig:knlc}.  
The singularity is located at the point of phase transition.\label{fig:knlr}}

In contrast to the above analysis, where the cosmological constant $\Lambda$
has been treated as a fixed background parameter, it is possible to raise $\Lambda$ 
at the level of an intrinsic parameter of the black hole and to consider it 
as an extensive thermodynamic variable \cite{cal00}. This is an interesting 
possibility that follows from the Kaluza-Klein reduction of certain supergravity 
theories which are relevant in M-theory \cite{dnp86}. In GTD this possibility can
easily be handled. In fact, in this case the phase space is 9-dimensional with
coordinates $Z^A = \{M,S,Q,J, \Lambda, T,\phi,\Omega, L\}$, where $L$ is 
the  intensive coordinate dual to $\Lambda$. The construction
of the Riemannian structure in ${\cal T}$ is straightforward, according to 
Eq.(\ref{giiup}). Furthermore, the coordinates of the space of equilibrium 
states can be chosen as $E^a = \{S,Q,J,\Lambda\}$ so that the corresponding
thermodynamic metric becomes
\be
g=(SM_S + QM_Q + JM_J+\Lambda M_\Lambda)\left(
\begin{array}{cccc}%[pos]
-M_{SS}& 0 & 0 & 0 \\
0 &  M_{QQ} &  M_{QJ} & M_{Q\Lambda} \\
0 &  M_{QJ} & M_{JJ} & M_{J\Lambda} \\
0 & M	_{Q\Lambda} & M_{J\Lambda} & M_{\Lambda\Lambda}
\end{array}
\right) \ .
\label{gknadsl}
\ee
The scalar curvature of this metric is again singular at the roots of $M_{SS}=0$
so that, in principle, it can reproduce the structure of the phase transitions 
of the KN-AdS black hole.

%%%%%%%%%%%%%%%%%%%%%%%%%%%%%%%%%%%%%%%%%%%%%%%%%%%%%%%%%%%%%%
%%%%%%%%%%%%%%%%%%%%%%%%%%%%%%%%%%%%%%%%%%%%%%%%%%%%%%%%%%%%%%
\section{Reissner-Nordstr\"om-AdS black hole in arbitrary dimensions}
\label{sec:highrn}

For a spacetime with arbitrary dimension $D$, the Einstein-Maxwell action 
with cosmological constant can be written as
\be
S_{_{EM}}= \frac{1}{16\pi} \int_{M^D} d^D x [-{\rm det}(g_{\mu\nu})]^{1/2}
\left[ R - F_{\mu\nu}F^{\mu\nu} +\frac{(D-1)(D-2)}{l^2}\right] \ ,
\ee
where $l$ is the characteristic length of the AdS background that determines
the cosmological constant by 
\be
\Lambda = - \frac{(D-1)(D-2)}{2l^2}\ .
\ee
Then the metric for the RN-AdS black hole may be written in static coordinates as 
\cite{jose99,cham99}
\be
ds^2 = -f(r) dt^2 + f^{-1}(r) dr^2 + r^2 d\Omega^2_{D-2} \ ,
\ee
where $d\Omega^2_{D-2}$ is the metric on the unit $(D-2)-$sphere. The function $f(r)$
can be expressed as
\be
f(r) = 1- \frac{\mu}{r^{D-3}} + \frac{q^2}{r^{2(D-3)}} + \frac{r^2}{l^2} \ ,
\ee
where $\mu$ and $q$ are intrinsic parameters related to the mass and charge of
the black hole. This is an exact solution of Einstein-Maxwell equations with 
electromagnetic potential
\be
A_t = - \left[\frac{D-2}{2(D-3)}\right]^{1/2}
\frac{q}{r^{D-3}} \ .
\ee
The outer horizon is situated at $r=r_+$ where $r_+$ is the largest root of 
the equation $f(r)=0$. From this algebraic equation it follows that
\be
\mu  = r_+^{D-3} + \frac{q^2}{r_+^{D-3}} + \frac{r_+^{D-1}}{l^2} \ .
\label{mu}
\ee
Furthermore, the horizon area is given by
\be
A = \omega_{_{D-2}} r_+^{D-2} \ ,
\ee
where $\omega_{_{D-2}} = 2\pi^{(D-1)/2}/\Gamma[(D-1)/2]$ is the volume 
of the unit $(D-2)-$sphere. The calculation of the physical mass and charge 
can be carried out either by using an appropriate generalization of the 
Arnowitt-Deser-Misner (ADM), which includes the case of asymptotically anti-de Sitter
spacetimes \cite{abb82,hawhor96}, or by using as a guide 
the laws of black hole thermodynamics \cite{gib05}. The resulting quantities can be
written as
\be
M=\frac{(D-2) \omega_{_{D-2}}}{16\pi}\, \mu \, \qquad
Q = \frac{[2(D-2)(D-3)]^{1/2} \omega_{_{D-2}} }{8\pi} \, q \ .
\ee
After some algebraic manipulations which involve the expressions given above 
for horizon area in the form $S=A/4$, the mass, charge, and the parameter $\mu$, 
we obtain
\be
M=\frac{(D-2)\omega_{_{D-2}}}{16\pi}
\left(\frac{4S}{\omega_{_{D-2}}} \right)^{\frac{D-1}{D-2}}
\left[ \frac{1}{l^2}+\left(\frac{\omega_{_{D-2}}}{4S} \right)^{\frac{2}{D-2}}
+\frac{2\pi^2 Q^2}{(D-2)(D-3) S^2}\right] \ .
\label{mrnads}
\ee
This is the fundamental equation for the RN-AdS black hole in arbitrary dimensions. In 
the special case $D=4$ we recover the expression obtained in the last section. From the
fundamental equation it is easy to derive all important thermodynamic variables. So we obtain the temperature 
\be
T=\frac{\partial M}{\partial S} = 
\frac{1}{4\pi}\left(\frac{4S}{\omega_{_{D-2}}} \right)^{\frac{1}{D-2}}
\left[ \frac{(D-1)}{l^2}+(D-3)\left(\frac{\omega_{_{D-2}}}{4S} \right)^{\frac{2}{D-2}}-\frac{2\pi^2 Q^2}{(D-2) S^2}\right]
 \ ,
\label{trnads}
\ee
the electric potential,
\be
\phi = \frac{\partial M}{\partial Q} = 
\frac{\pi Q}{(D-3) S}\left(\frac{4S}{\omega_{_{D-2}}} \right)^{\frac{1}{D-2}}
\ ,
\label{phirnads}
\ee
and the heat capacity
\be
C= T \frac{\partial S}{\partial T} = \frac{M_S}{M_{SS}} = 
\frac{(D-2)S\left[ \frac{D-1}{l^2}+(D-3)\left(\frac{\omega_{_{D-2}}}{4S} \right)^{\frac{2}{D-2}}-\frac{2\pi^2 Q^2}{(D-2) S^2}\right] }
{ \frac{D-1}{l^2}-(D-3)\left(\frac{\omega_{_{D-2}}}{4S} \right)^{\frac{2}{D-2}}+\frac{2(2D-5)\pi^2 Q^2}{(D-2) S^2} } \ .
\ee
These thermodynamic variables are then considered as independent quantities at the
level of the 5-dimensional thermodynamic 
phase space ${\cal T}$ which can be coordinatized by 
$Z^A = \{M,S,Q,T,\phi\}$. The Riemannian structure of ${\cal T}$ is determined in 
this case by the metric
\be
G = (dM - T dS - \phi dQ)^2 + (TS + \phi Q) \left( - dT d S + d\phi dQ\right)\ .
\label{guprnads}
\ee
The space of equilibrium states ${\cal E}$ can be chosen as being determined by
the simple mapping $\varphi: \{S, Q\} \mapsto \{M(S,Q), S, Q, T(S,Q), \phi(S,Q)\}$,
where the explicit dependence of the intensive variables is as given above. The 
thermodynamic metric on ${\cal E}$ can be computed by means of the pullback 
$g=\varphi^*(G)$ that yields
\be
g = (SM_S + Q M_Q) \left(  
\begin{array}{cc}%[pos]
-M_{SS}& 0  \\
0 &  M_{QQ}  
\end{array}
\right) \ .
\label{grnads}
\ee
As for the corresponding scalar curvature, we have performed an exhaustive 
analysis of the singularities and obtained a behavior similar to that of
the heat capacity. Figures \ref{fig:rnlc} and \ref{fig:rnlr} 
show a characteristic example of the singular behavior of the heat
capacity and the thermodynamic curvature. Our analysis shows 
that the curvature singularities reproduce the structure of the phase 
transitions of the Reissner-Nordstr\"om black hole in arbitrary dimensions.
\DOUBLEFIGURE{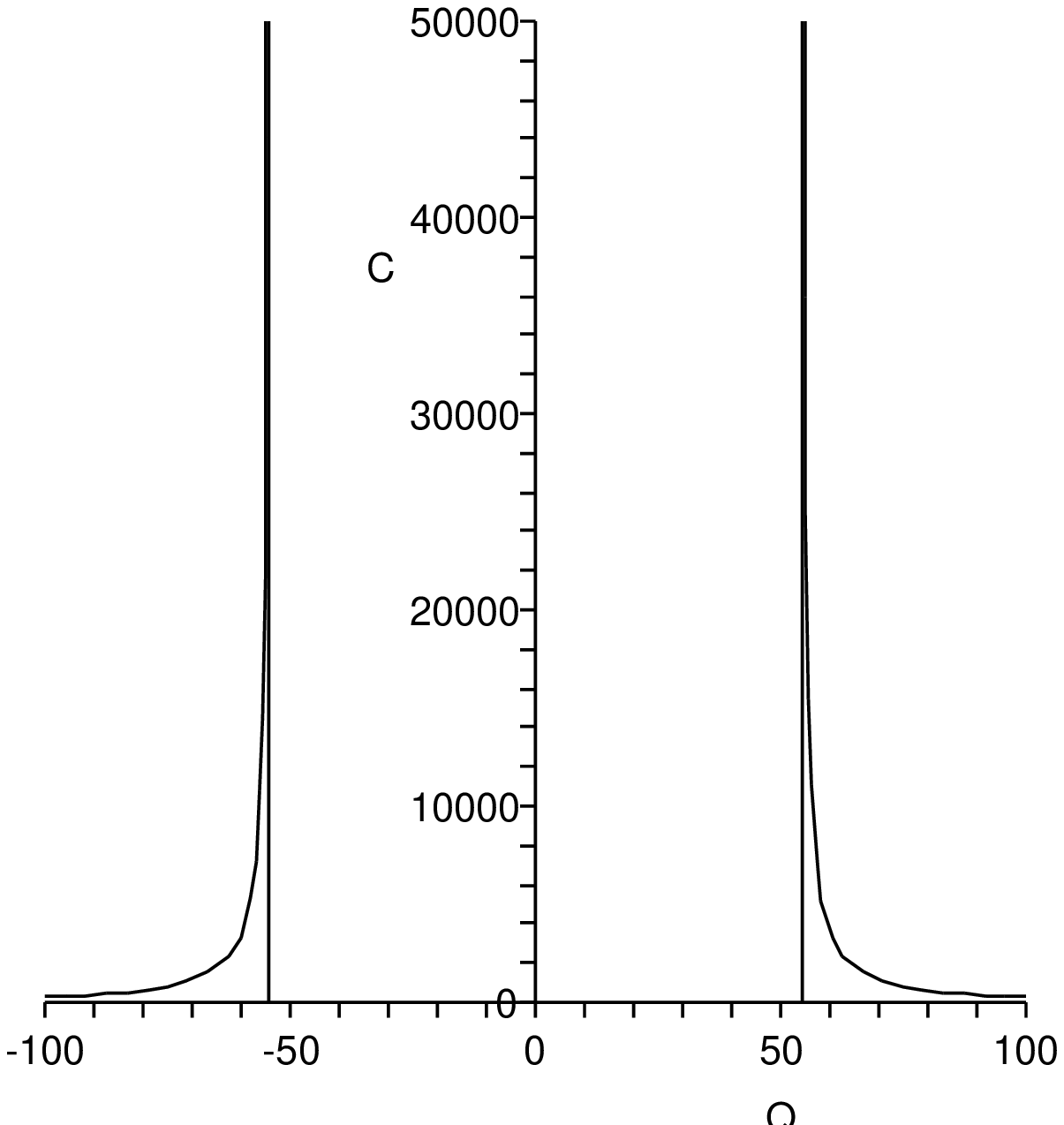,width=7cm}{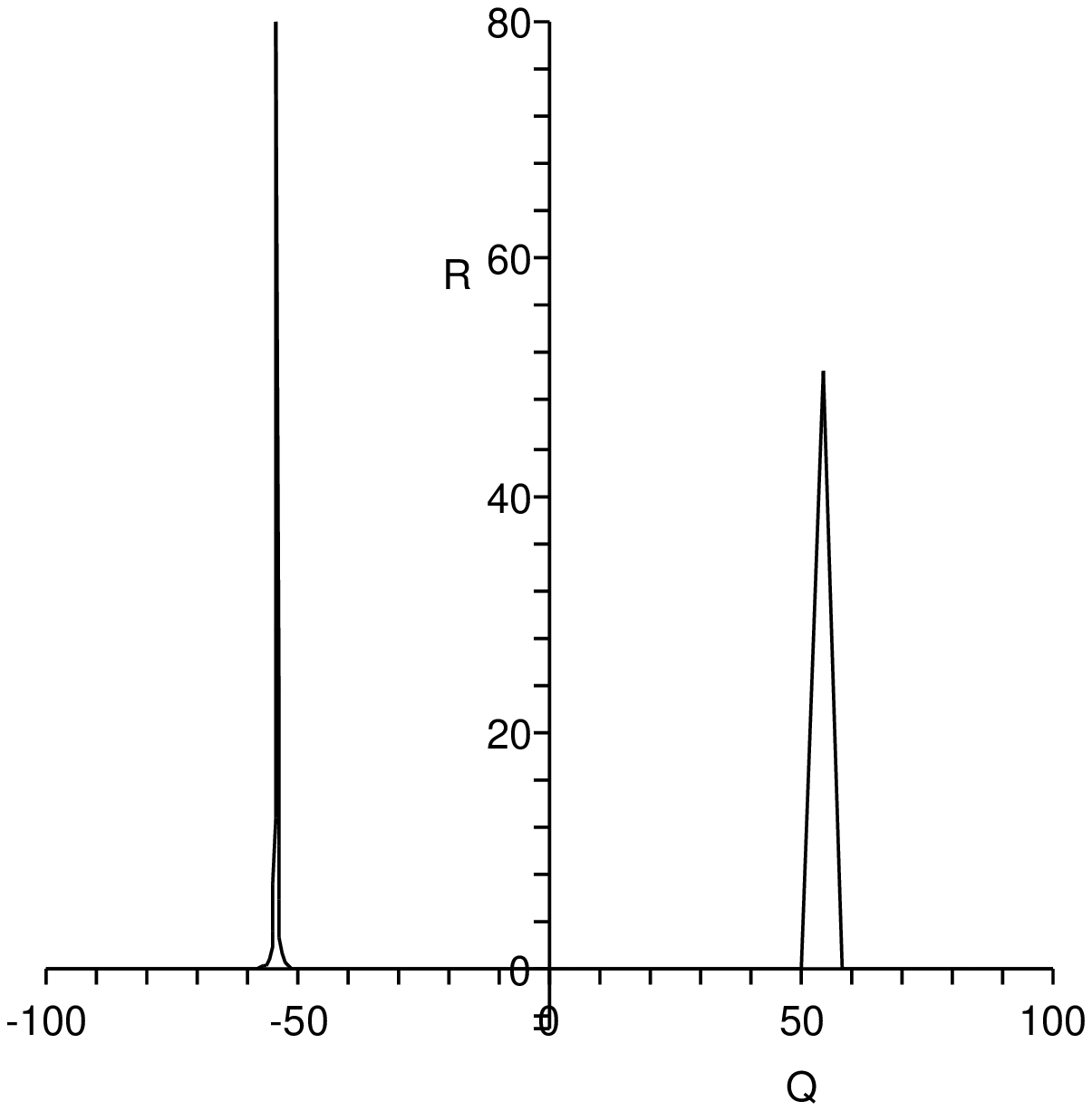,width=7cm}
{The heat capacity $C$ as a function of the charge $Q$ for the 10-dimensional 
RN-AdS black hole. Here $\Lambda=-1$ and $S=100$. Phase transitions are 
clearly seen as divergencies of $C$.\protect\label{fig:rnlc}}
{The scalar curvature $R$ for the thermodynamic metric 
(\protect\ref{grnads}) as a function of the charge $Q$.  
The values of the remaining parameters are as in figure \protect\ref{fig:rnlc}.  
The locations of curvature singularities and phase transitions coincide. 
\protect\label{fig:rnlr}
}

%%%%%%%%%%%%%%%%%%%%%%%%%%%%%%%%%%%%%%%%%%%%%%%%%%%%%
%%%%%%%%%%%%%%%%%%%%%%%%%%%%%%%%%%%%%%%%%%%%%%%%%%%%%
\section{Kerr-AdS black hole in arbitrary dimensions}
\label{sec:highkerr}

In an arbitrary spacetime of dimension $D$, the Kerr-anti 
de Sitter metric is an exact solution
to the equations $R_{\mu\nu} + (D-1) l^{-2}g_{\mu\nu}=0$, which in Boyer-Lindquist
coordinates can be expressed as \cite{kerrall}
\bea
& ds^2 & =  -  
W \left(1 + \frac{r^2}{l^2}\right) dt^2 + \frac{2m}{U}\left(W dt - \sum_{i=1}^N \frac{a_i\mu_i^2}{\Xi_i}d\phi_i\right)^2 
+\sum_{i=1}^N \frac{r^2+a_i^2}{\Xi_i}\mu_i^2 d\phi_i^2 \nonumber\\
& + & \frac{U}{V-2m} dr^2 + \sum_{i=1}^{N+\epsilon} \frac{r^2+a_i^2}{\Xi_i} d\mu_i^2 -
\frac{1}{l^2W}\left(1 + \frac{r^2}{l^2}\right)^{-1} \left(\sum_{i=1}^{N+\epsilon} \frac{r^2+a_i^2}{\Xi_i}\mu_i d\mu_i\right)^2 ,
\eea
where $N=[(D-1)/2]$ is the maximum number of independent rotational parameters $a_i$ in $N$ independent, orthogonal 2-planes, $m$ is a parameter related to the total mass of the
black hole,
the parameter $\epsilon$ is defined as $\epsilon= (D-1)\ {\rm mod}\ 2$, $t$ is the time coordinate,  $r$ is the radial 
coordinate, $\phi_i$ are $N$ different azimuthal angles, and $\mu_i$ are $(N+\epsilon)$ direction cosines. Moreover, the functions
appearing in the metric are
\be
W = \sum_{i=1}^{N+\epsilon} \frac{\mu_i^2}{\Xi_i} \ , \  U = r^{\epsilon}\sum_{i=1}^{N+\epsilon} \frac{\mu_i^2}{r^2+a_i^2}
\prod_{j=1}^N (r^2 + a_j^2)  \ ,
\ee
\be 
V = r^{\epsilon - 2} \left( 1 + \frac{r^2}{l^2}\right) \prod_{i=1}^N (r^2 + a_i^2) \ , \
\Xi_i = 1 - \frac{a_i^2}{l^2}\ .
\ee
The outer horizon is situated at the radial distance $r=r_+$, where $r_+$ is 
the largest positive root of the algebraic equation
\be
r_+^{\epsilon - 2} \left( 1 + \frac{r_+^2}{l^2}\right) \prod_{i=1}^N (r_+^2 + a_i^2) - 
2m =0 \ ,
\label{outerh}
\ee
which once solved can be used to find explicitly the value of the entropy $S=A/4$. 
As in the previous cases, the presence of the cosmological constant requires especial 
care for the determination of the physical parameters. Let us denote by $M$ the physical 
mass of the black hole and by $J_i$ the momentum corresponding to the angular 
velocity $\Omega_i$, measured by a non-rotating observer at infinity. Using as a
guide the laws of thermodynamics, it can be shown that \cite{gib05}
\be
M= \frac{m \omega_{_{D-2}}}{4\pi} \left(\prod_{i=1}^N \Xi_i\right)^{-1}\ \left(
\sum_{j=1}^N \frac{1}{\Xi_j} - \frac{1}{2}\right)\ , \quad
S=\frac{\omega_{_{D-2}} }{4 r_+} \prod_{i=1}^N \frac{r_+^2 + a_i^2}{\Xi_i} \ , 
\qquad {\rm for\ odd\ } D 
\label{massodd}
\ ,
\ee
\be
M= \frac{m \omega_{_{D-2}}}{4\pi} \left(\prod_{i=1}^N \Xi_i\right)^{-1}\ \sum_{j=1}^N 
\frac{1}{\Xi_j}\ , \quad
S=\frac{ \omega_{_{D-2}} }{4} \prod_{i=1}^N \frac{r_+^2 + a_i^2}{\Xi_i} \ , 
 \qquad {\rm for\ even\ } D 
\ , 
\label{masseven}
\ee
\be
J_i = \frac{ma_i\omega_{_{D-2}}}{4\pi \Xi_i} \left(\prod_{j=1}^N \Xi_j\right)^{-1}\ , 
\qquad {\rm for\ arbitrary\ } D\ .
\label{angmom}
\ee
The next step is the derivation of a Smarr-like formula which relates all the physical 
parameters entering the metric, i.e. $M=M(S,J_i)$ or equivalently $S=S(M,J_i)$. To this
end it is necessary to express the outer radius $r_+$ in terms of the physical parameters.
In case this is possible, we first select a representation for the construction of GTD. 
Let us consider the $M-$representation. Then, the extensive thermodynamic variables are
$S$ and $J_i$. If we denote by $T$ and $\Omega_i$ the corresponding dual intensive 
thermodynamic variables, the coordinates of the phase space ${\cal T}$ can be chosen
as $Z^A=\{M,S,J_i,T,\Omega_i\}$ so that dim$({\cal T})=3 +2[(D-1)/2]$, when the maximum 
number of angular momenta is considered. The Riemannian structure of ${\cal T}$ is 
completed with the metric 
\be
G=\left(dM - TdS - \sum_{i=1}^N \Omega_i d J_i\right)^2 + \left(
ST + \sum_{i=1}^N \Omega_i J_i\right)
\left(- dS dT + \sum_{i=1}^N d \Omega_i d J_i\right) \ .
\ee
The Riemannian submanifold of equilibrium states ${\cal E}$ is defined by means 
of the smooth mapping $
\varphi: \{S,J_i\} \longmapsto \{ M(S,J_i), S, J_i, T(S,J_i), \Omega_i(S,J_i)\}$,
which induces the first law of thermodynamics 
$dM = TdS + \sum_{i=1}^N \Omega_i d J_i$
and the $(N+1)-$dimensional metric 
\be
g=\left(SM_S + \sum_{i=1}^N J_i M_{J_i} \right)\left(
\begin{array}{cccc}%[pos]
-M_{SS}&  0           & 0     & 0 \\
     0 &  M_{J_1 J_1} & ...  & M_{J_1J_N}  \\
     0 & .            & ... & . \\
     0 & M_{J_N J_1}  & ... & M_{J_N J_N}
\end{array}
\right) \ .
\label{gkerrads}
\ee
It is straightforward to see that the curvature scalar of this thermodynamic metric 
contains the term 
\be
\left(SM_S + \sum_{i=1}^N J_i M_{J_i} \right)^N [{\rm det}(M_{ij})]^N M_{SS}^2 
\label{denokerrads}
\ee
in its denominator, where $M_{ij} = \partial^2M/(\partial J_i \partial J_j)$, 
so that 
curvature singularities could appear at the points where the condition $M_{SS}=0$ is
satisfied. On the other hand, as mentioned above, in the mass representation the 
heat capacity can be expressed as $C= M_S/M_{SS}$. It follows then that phase
transitions that occur when the heat capacity diverges, could be represented 
as curvature singularities of the thermodynamic metric. This holds, of course,
only if the expressions appearing in the numerator of the curvature do not 
eliminate the zeros of $M_{SS}$. We note that in (\ref{denokerrads}) the term
in round brackets cannot be zero because it can be shown to be 
proportional to the total mass $M$ as a result of Euler's identity. As for the
determinant of $M_{ij}$, its zeros, if any, can be found only if the fundamental
equation $M=M(S,J_i)$ can be written explicitly.

The determination of the fundamental equation turns out to be a non-trivial problem.
In fact, one way to determine it is to compute the solutions of the algebraic
equation (\ref{outerh}) which, in general, is a polynomial of order $D$ in $r_+$.
We were not able to find explicit solutions. However, the case of vanishing 
cosmological constant ($l^2\rightarrow \infty)$ and only one non-vanishing
rotational parameter, say, $a_1=a$, can be manipulated explicitly. 
From Eqs.(\ref{outerh})-- (\ref{angmom}), we obtain for this specific case
\be
(r_+^2 + a^2) r_+^{D-5} - 2m = 0 \ ,
\ee 
\be
M=\frac{m\omega_{_{D-2}}} {4\pi}(D/2-1)\ ,\ \
S= \frac{ \omega_{_{D-2}} }{4} (r_+^2 + a^2) r_+^{D-4}\ ,\ \
J=\frac{ma\omega_{_{D-2}}} {4\pi}\ .
\ee
A straightforward manipulation of these equations results in the expression
\be
M= \frac{D/2-1}{\pi} \left(\frac{ \omega_{_{D-2}} }{2^D}\right)^{\frac{1}{D-2}}
S ^{ \frac{D-3}{D-2}} \left( 1 + 4\pi^2 \frac{J^2}{S^2}\right)^{ \frac{1}{D-2}} \ ,
\ee
which constitutes the corresponding fundamental equation.  In turn, 
the heat capacity can be expressed as
\be
C=-\frac{(D-2)S
\left[ 3S^2+20\pi^2 J^2-D(S^2+4\pi^2 J^2)\right]\left( S^2+4\pi^2 J^2\right)}
{3S^4+24\pi^2 J^2 S^2+240\pi^4 J^4-D(S^4+48\pi^4 J^4)}\ . 
\label{heatkerr}
\ee
The corresponding thermodynamic metric of the space of equilibrium states
reduces to the 2-dimensional metric
\be
g = (SM_S + JM_J)\left(- M_{SS} d S^2 + M_{JJ} d J^2\right) \ ,
\label{gkerrddown}
\ee
independently of the dimension $D$. The expression for the scalar curvature 
cannot be transformed into a compact form. Only the denominator can be shown
to contain the expression 
\be
(D-3)[3S^4+24\pi^2 J^2 S^2+240\pi^4 J^4-D(S^4+48\pi^4 J^4)]^2 \ ,
\ee
which determines the phase transition structure of the heat capacity (\ref{heatkerr}).
To see that the singular behavior of the scalar curvature coincides with that of
the heat capacity we perform a detailed numerical analysis of both expressions. 
The characteristic singular behavior is depicted in figures \ref{fig:kerrdc} and
\ref{fig:kerrdr}. For all analyzed regions
a similar behavior was detected, showing that in fact the points where phase transitions
occur are characterized by curvature singularities of the thermodynamic metric.
\DOUBLEFIGURE{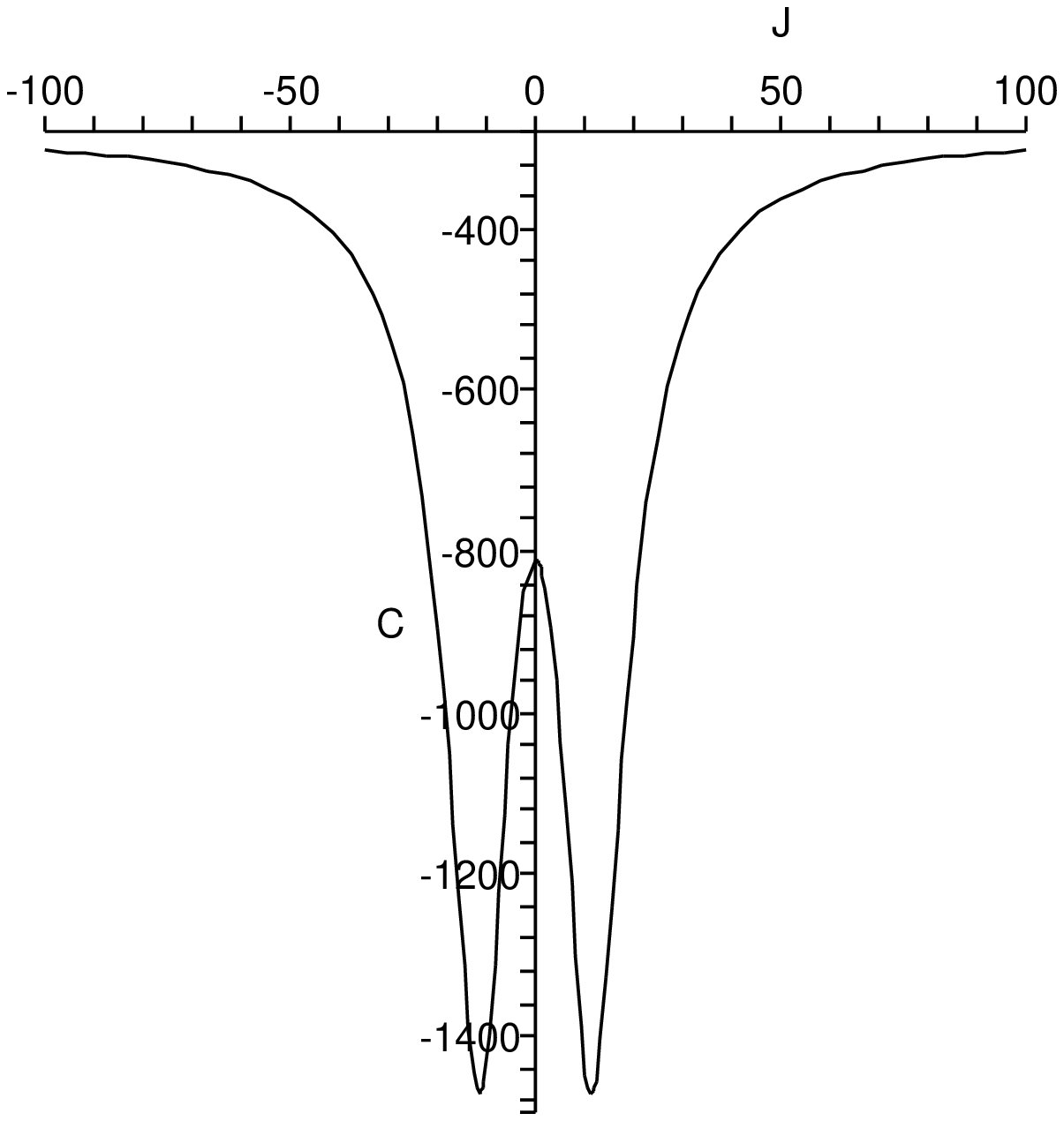,width=7cm}{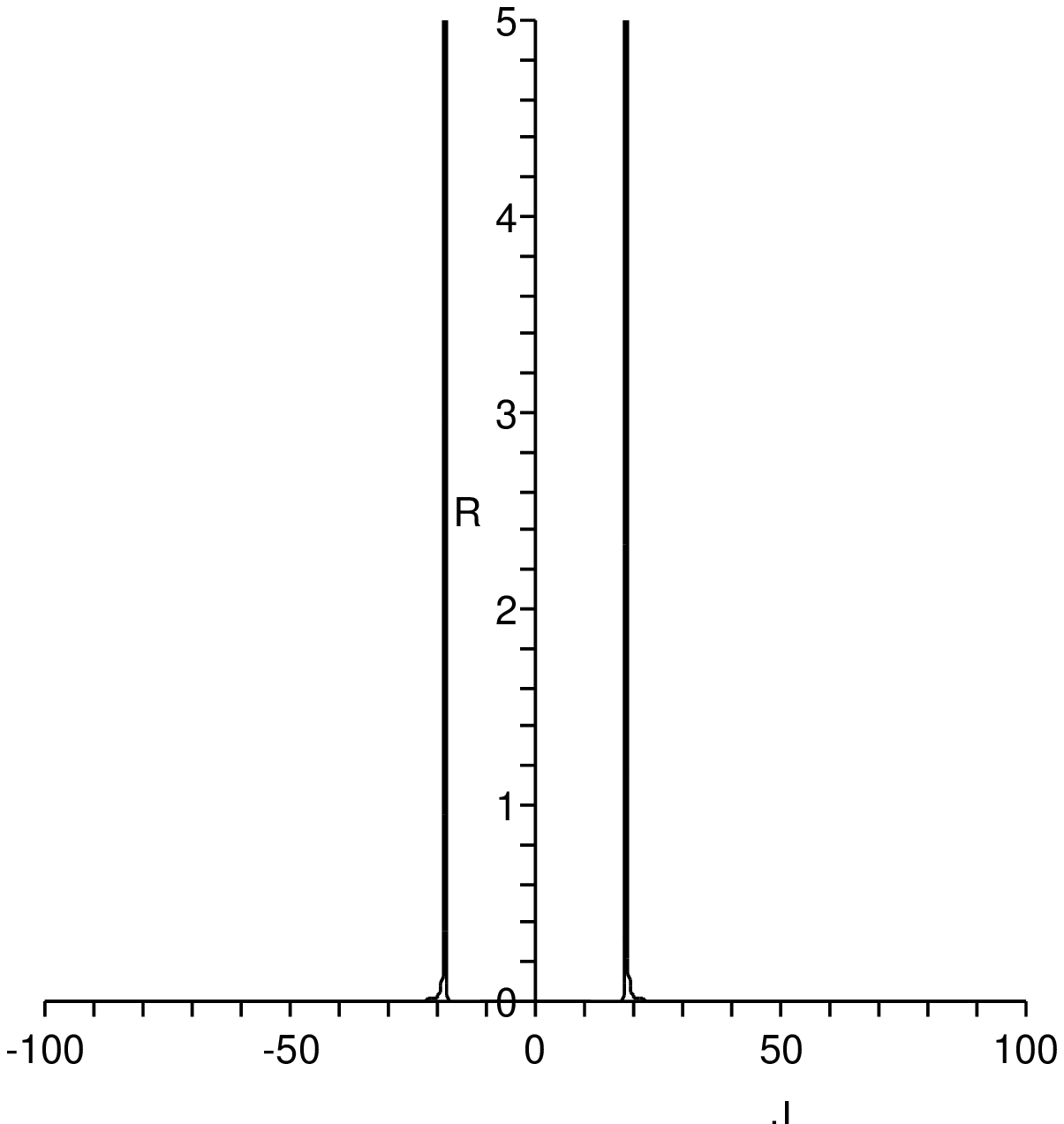,width=7cm}
{The heat capacity $C$ as a function of the angular momentum $J$ for 
the 10-dimensional Kerr black with one rotational parameter. 
Here $\Lambda=0$ and $S=100$. \protect\label{fig:kerrdc}}
{The scalar curvature $R$ for the thermodynamic metric 
(\protect\ref{gkerrddown}) as a function of the angular momentum $J$.
The remaining parameters are chosen as in figure \protect\ref{fig:kerrdc}.  
\protect\label{fig:kerrdr}
}    
%%%%%%%%%%%%%%%%%%%%%%%%%%%%%%%%%%%%%%%%%%%%%%%%%%%%%%%%%%%%
\section{The role of statistical ensembles}
\label{sec:ensembles}
In the above applications of GTD of black holes, the starting point of the analysis
is the fundamental thermodynamic equation. Nevertheless, it is known that the thermodynamic 
properties of black holes can drastically depend on the choice of statistical ensemble
\cite{cal00,jose99,cham99,coreanos08}. The fact that different ensembles lead to different 
heat capacities implies that the phase transitions structure depends on the statistical
model under consideration. The question arises whether GTD is able to correctly handle 
this dependence. We will show in this section that the answer is in the affirmative. 
For the sake of concreteness and simplicity we will show this in the explicit example of
the RN-AdS black hole in $D=4$. 

Let us first consider the canonical ensemble which is characterized by a fixed charge. 
In this case, according to Eqs.(\ref{mrnads})--(\ref{phirnads}), 
the relevant thermodynamic variables can be expressed as 
\be
M = \frac{r_+}{2}\left(1+\frac{Q^2}{r_+^2}+\frac{r_+^2}{l^2}\right)\ , \quad
T = \frac{1}{4\pi}\left(\frac{1}{r_+}-\frac{Q^2}{r_+^3}+ \frac{3r_+^2}{l^2}\right)\ ,
\quad \phi= \frac{Q}{r_+}\ .
\label{fixedq}
\ee
where we have used the relation $S=\pi r_+^2$. Then, the heat capacity 
$C_Q = (\partial M / \partial T)_Q$ becomes
\be
C_Q = 2\pi r_+^2\left[\frac{3r_+^4+l^2(r_+^2-Q^2)}{3r_+^4-l^2(r_+^2-3Q^2)}\right]\ .
\label{cqrnads4}
\ee

The analysis of this case in the context of GTD can be carried out as in Section 
{\ref{sec:highrn}. The metric $G$ of the phase space ${\cal T}$ 
is given by (\ref{guprnads}), 
whereas the metric $g$ of the space of equilibrium states ${\cal E}$
coincides with (\ref{grnads}) 
and can be written explicitly as
\bea
g & = & \left(SM_S+QM_Q\right) \left( -M_{SS} dS^2 + M_{QQ} dQ^2\right)\nonumber \\
& = & \frac{3r_+^4+l^2(3Q^2+r_+^2)}{4l^2r_+^2}\left[
-\frac{3r_+^4+l^2(3Q^2-r_+^2)}{8\pi^2l^2r_+^4}\, dS^2 + dQ^2\right]\ .
\eea
This metric determines the geometric properties of the manifold ${\cal E}$. In 
particular, the curvature scalar can be expressed in the compact form
\be
R = - \frac{48l^4r_+^4{\cal N}(r_+,Q)}{[3r_+^4+l^2(r_+^2+3Q^2)]^3
[3r_+^4-l^2(r_+^2-3Q^2)]^2 }\ ,
\ee
where 
\be
{\cal N}(r_+,Q) = 3l^4Q^4(2l^2-21r_+^2) - 3l^2r_+^2 Q^2(2l^4-5l^2r_+^2+6r_+^4)
-r_+^6(2l^4+3l^2r_+^2-45r_+^4) .
\ee
The curvature singularities are then given by the zeros of the equation
\be
3r_+^4-l^2(r_+^2-3Q^2) = 0
\ee
which are also the blow-up points of the heat capacity (\ref{cqrnads4}). This shows
that in the canonical ensemble approach the structure of the phase transitions
coincides with the singular structure of the thermodynamic curvature.

We now consider the grand canonical ensemble that corresponds to a fixed electric
potential $\phi$. To obtain the grand canonical potential one can use the Euclidean 
action method as in \cite{cham99} or, equivalently, apply a  
Legendre transformation as in \cite{coreanos08} which interchanges the role of the
variables $Q$ and $\phi$. That is to say, one introduces a new thermodynamic 
potential $\tilde M$ by means of the Legendre transformation
\be
\tilde M = M - \phi Q \ .
\ee
In \cite{coreanos08}, the potential $\tilde M$ is interpreted as  the internal energy. 
Then, from Eq.(\ref{fixedq}) we obtain 
\be
\tilde M = \frac{r_+}{2}\left(1-\phi^2 +\frac{r_+^2}{l^2}\right) \ .
\ee
On the other hand, calculating the total differential of $\tilde M$ and 
using the first law of thermodynamics, we obtain $d\tilde M = T dS - Qd\phi$ 
so that the dual thermodynamic variables 
$T=\partial \tilde M/\partial S$ and 
$Q=-\partial \tilde M/\partial \phi $ can be written as
\be
T= \frac{1}{4\pi r_+}\left( 1 - \phi^2 + \frac{3r_+^2}{l^2}\right)\ ,
\quad Q = \phi r_+ \ .
\ee
Furthermore, it is straightforward to calculate the heat capacity
%\footnote{Notice that this expression for the heat capacity $C_\phi$
% of the grand canonical ensemble cannot be obtained by simply substituting 
% $Q=\phi r_+$ in the expression for $C_Q$. 
% This explains the slight difference between our result and that of \cite{coreanos08}.}  
\be
C_\phi= \left(\frac{\partial \tilde M}{\partial T}\right)_\phi 
=  2\pi r_+^2\left[\frac{3r_+^2+l^2(1-\phi^2)} {3r_+^2-l^2(1-\phi^2)} \right]\ .
\ee
Hence, the blow-up points that determine the phase transitions structure coincide with the 
solutions of the equation
\be 
3r_+^2-l^2(1-\phi^2) = 0 \ .
\label{zerosphi}
\ee

We now analyze the case of the grand canonical ensemble in GTD. The 
new thermodynamic potential is $\tilde M = \tilde M (S,\phi)$ so that the 
coordinates in the phase space ${\cal T}$ are $Z^A=\{\tilde M ,E^a, I^a\}
= \{\tilde M, S, \phi, T, - Q\}$. Then, the fundamental contact form  
can be written it its canonical form as $\Theta = d\tilde M - T d S + Q d\phi$. 
The particular metric (\ref{giiup}) we are using here for the geometry of 
the manifold ${\cal T}$ takes the form
\be
G=  \left(d\tilde M - T d S + Q d\phi\right)^2 - (ST-\phi Q)(dSdT + d\phi d Q)
\ .
\ee
The submanifold ${\cal E}$ of equilibrium states with coordinates $E^a=\{S,\phi\}$
is defined by means of the smooth map $\varphi: {\cal E} \longrightarrow {\cal T}$ 
that in this case implies the explicit dependence 
\be
\varphi: \{E^a\} \longmapsto \{Z^A(E^a)\} = \left\{\tilde M (S,\phi), S, \phi, 
T=\frac{\partial\tilde M}{\partial S}, 
-Q =  \frac{\partial\tilde M}{\partial \phi}\right\}\ .
\ee
In turn, the geometric properties of ${\cal E}$ are described by the metric 
$g=\varphi^*(G)$ which becomes
\bea
g&=& \left(S\tilde M _S + \phi \tilde M _\phi\right) 
\left(-\tilde M_{SS} d S^2 + \tilde M _{\phi\phi} d\phi^2\right)\nonumber\\
& = & - \frac{3r_+^2-l^2(5\phi^2-1)}{4l^2}\left[
\frac{3r_+^2-l^2(1-\phi^2)}{8\pi^2l^2r_+^2}\, dS^2 + r_+^2\, d\phi^2\right] 
\eea
A straightforward calculation of the scalar curvature for this metric yields
\be
R= - \frac{ 16l^4{\cal N}(r_+,\phi)}{r_+^2
[3r_+^2-l^2(5\phi^2-1)]^3 [3r_+^2-l^2(1-\phi^2)]^2}\ ,
\ee
where 
\be
 {\cal N}(r_+,\phi)= 5l^4\phi^4(l^2+24r_+^2) +3l^2r_+^2\phi^2(60r_+^2-47l^2) 
 +3(l^6+l^4r_+^2-6l^2r_+^4+36r_+^6) \ .
\ee
From the denominator of this expression and Eq.(\ref{zerosphi}), 
we conclude that there exist curvature singularities at those points where 
the heat capacity $C_\phi$ blows up. There is an additional term in the 
denominator which under the condition $3r_+^2-l^2(5\phi^2-1)=0$ can, in principle,
lead to curvature singularities. However, it can easily be seen that this
term is proportional to the conformal factor of the metric, i. e., 
$S\tilde M _S + \phi \tilde M _\phi$, which in turn, according
to Euler's identity, is proportional to the internal energy $\tilde M$ of the 
black hole. This seems to prohibit the appearance of additional singularities.

We conclude that GTD can handle correctly the different thermodynamic schemes 
that follow from different statistical ensembles for the RN-AdS black hole 
in $D=4$. Similar analysis can be performed for other types of black holes 
and we expect similar results. The fact that in general the denominator of 
the scalar curvature contains the denominator of the heat capacity is an 
indication that, even in the case of different statistical ensembles, there exists
a correspondence between singular points
of the heat capacity and curvature singularities.

%%%%%%%%%%%%%%%%%%%%%%%%%%%%%%%%%%%%%%%%%%%%%%%%%%%%%
%%%%%%%%%%%%%%%%%%%%%%%%%%%%%%%%%%%%%%%%%%%%%%%%%%%%%
\section{Discussion and conclusions}
\label{sec:con}

The main result of this work is that in the space of equilibrium states 
of all known asymptotically anti-de Sitter black holes in arbitrary dimensions 
there exists a thermodynamic metric whose curvature is singular at those
points where phase transitions of the heat capacity occur. This has been
shown by considering a particular metric in the thermodynamic phase space,
and applying the formalism of geometrothermodynamics. An important property
of our choice of thermodynamic metric is that it is invariant with respect
to Legendre transformations so that the properties of our geometric
description of thermodynamics are independent of the choice of thermodynamic
potential and representation. 

The explicit examples considered in this work 
include the Reissner-Nordstr\"om black holes and the Kerr black holes on
an anti-de Sitter background in arbitrary dimensions ($D\geq 4$). 
In the last example, we were able to explicitly analyze
only the case of one rotational parameter with vanishing cosmological constant.
The most general case of non-vanishing cosmological constant and arbitrary 
number of rotational parameters leads to a set of algebraic equations which 
we were not able to solve. Hence the fundamental thermodynamic equation
cannot be written explicitly. Nevertheless, even in this general case 
our results show that if the numerator of the resulting expression for
the curvature scalar does not cancel the zeros of the denominator, then 
the curvature singularities are situated at those points where the 
heat capacity diverges, a fact that  in ordinary thermodynamics is considered 
as an indication of a phase transition. The four-dimensional 
Kerr-Newman black hole on an anti-de Sitter background was also analyzed, obtaining
similar results. The Kerr-Newman-AdS black hole is not known in higher dimensions.
Thus, the examples considered in this work include all known higher 
dimensional solutions for black holes on an anti-de Sitter background.

The starting point in our geometrothermodynamical analysis is the metric $G$ of
the thermodynamic 
phase space. In this work we used the particular metric (\ref{giiup}) which 
was chosen such that the metric $g$ of the space of equilibrium 
states takes the specific form (\ref{giidown}), whose determinant becomes
proportional to $\partial^2\Phi/\partial E^1 \partial E^1$. In fact, the use 
of the metric $\eta_{ab}$, instead of $\delta_{ab}$, in $G$ guarantees that all 
non-diagonal terms of the form  $\partial^2\Phi/\partial E^1 \partial E^k$, with 
$k\neq 1$, vanish. 
Furthermore, it is
known that the scalar curvature always contains terms with the determinant of
the metric in their denominator. It can therefore be expected that 
there exist true curvature singularities at those points where 
 $\partial^2\Phi/\partial E^1 \partial E^1=0$. We use in this work the
$M-$representation, in which $\Phi=M$, and choose $E^1=S$ so that the singularities 
are expected at $\partial^2 M/\partial S ^2 = M_{SS}=0$. 
On the other hand, in this representation the 
heat capacity can be written as $C=M_S/M_{SS}$ with divergencies at 
$M_{SS}=0$. For all known asymptotically anti-de Sitter black holes we have
shown that the remaining terms of the thermodynamic scalar curvature 
do not cancel the zeros of $M_{SS}$. 
Thus  we conclude that the curvature of the
space of thermodynamic equilibrium states can be interpreted in an invariant 
manner as a measure of the thermodynamic interaction. 
This is in contrast 
with other studies \cite{aman06,cai07} where the curvature of the space of equilibrium 
states depends on the choice of a specific representation so that, for instance,
a flat thermodynamic metric can be associated to a system with non-trivial 
thermodynamic interaction.
  
In the case of the RN-AdS black hole in $D=4$, we analyzed the different thermodynamics
which follow from the choice of different statistical ensembles. In particular,
we considered the canonical ensemble, with fixed charge $Q$, 
and the grand canonical ensemble, with fixed electric potential $\phi$. 
We obtained the corresponding heat capacities $C_Q$ and $C_\phi$, 
which lead to different phase transitions structures. According
to our general procedure of GTD for black holes, we constructed 
explicitly for both ensembles the corresponding phase manifolds and 
the manifolds of equilibrium states. In both cases we obtained that
the there exists a correspondence between singular points
of the heat capacity and curvature singularities of the thermodynamic 
metric. We expect similar results in the analysis of more general black 
holes. The fact that in general the denominator of 
the scalar curvature contains the denominator of the heat capacity is an 
indication that GTD can handle correctly the different thermodynamic schemes 
that follow from different statistical ensembles.  

Legendre invariance is an important element of our approach. It limits
the number of metrics that can be used to describe ordinary thermodynamics
in terms of geometric concepts. It is also essential in order to obtain 
results that are independent of the choice of extensive variables and
thermodynamic potential. A different point of view, in which for a given
thermodynamic system there exists a preferred thermodynamic potential 
\cite{ruppmail,med08}, is necessary in order to explain the vanishing of Ruppeiner's 
thermodynamic curvature in cases where thermodynamic interaction is present as, 
for instance, in Reissner-Nordstr\"om black holes.  We believe that 
ordinary thermodynamics, which is Legendre invariant,
must not be changed when one tries to represent it in terms of 
geometric concepts.

The computer algebra system REDUCE 3.8 was used for most of the calculations 
reported in this work.

\section*{Acknowledgements}

This work was supported in part by Conacyt, Mexico, grant 48601.

% \appendix

% \section{}
% \label{}

%\appendix

%%%%%%%%%%%%%%%%%%%%%%%%%%%%%%%%%%%%%%%%%%%%%%%%%%%%%%%%%%%%%%%%%%%%% 
%%%%%%%%%%%%%%%%%%%%%%%%%%%%%%%%%%%%%%%%%%%%%%%%%%%%%%%%%%%%%%%%%%%%%%


\begin{thebibliography}{99}

\bibitem{frankel} T. Frankel, {\it  The Geometry of Physics: An Introduction}
(Cambridge University Press, Cambridge, UK, 1997).


\bibitem{her} R. Hermann, {\it Geometry, physics and systems} (Marcel 
Dekker, New York, 1973). 
 
\bibitem{wei1} F. Weinhold, {\it Metric Geometry of equilibrium thermodynamics I, II,
III, IV, V}, J. Chem. Phys. {\bf 63}, 2479, 2484, 2488, 2496 (1975); {\bf 65}, 558  (1976).
 
 
\bibitem{rup79} G. Ruppeiner, {\it Thermodynamics: A Riemannian geometric model},
Phys. Rev. A {\bf 20},  1608 (1979).


\bibitem{sal83} P. Salamon, E. Ihrig and R. S. Berry, {\it A group of coordinate
transformations which preserve the metric of Weinhold}, J. Math. Phys.  {\bf 24},  2515
(1983).


\bibitem{mru90} R. Mrugala, J. D. Nulton, J. C. Sch\"on, and P. Salamon, {\it
Statistical approach to the geometric structure of thermodynamics},
Phys. Rev. A {\bf 41},  3156 (1990). 

\bibitem{quev08} H. Quevedo, {\it Geometrothermodynamics of black holes}, Gen. Rel. Grav.  {\bf 40}, 971 (2008).

\bibitem{quev07} H. Quevedo, {\it Geometrothermodynamics}, J. Math. Phys. 
{\bf 48}, 013506 (2007). 


\bibitem{aqs08} J. L. \'Alvarez, H. Quevedo, and A. S\'anchez, {\it Unified geometric 
description of black hole thermodynamics}, Phys. Rev. D {\bf 77}, 084004 (2008). 



\bibitem{carter} B. Carter, {\it Hamilton-Jacobi and Schr\"odinger separable solutions
of Einstein's equations}, Commun. Math. Phys. {\bf 10}, 280 (1968). 

 

\bibitem{haw99} S. W. Hawking, C. J. Hunter, and M. M. Taylor-Robinson, {\it Rotating and the AdS/CFT correspondence},
Phys. Rev. D {\bf 59}, 064005 (1999).


\bibitem{rnall} X. Dianyan, {\it Exact solutions of Einstein and Einstein-Maxwell 
equations in higher-dimensional spacetime}, Class. Quantum Grav. {\bf 5},  871 (1988).
   

\bibitem{kerrall} G. W. Gibbons, H. L\"u, D. N. Page, and C. N. Pope, {\it The general Kerr-de Sitter metric in all dimensions}, J. Geom. Phys. {\bf 53}, 49 (2005).  


\bibitem{cal00} M. M. Caldarelli, G. Cognola, and D. Klemm, 
{\it Thermodynamics of Kerr-Newman-AdS black holes
and conformal field theories},  Class. Quantum Grav. {\bf 17}, 399 (2000). 

\bibitem{gib05} G. W. Gibbons, M. J. Perry, and C. N. Pope, {\it The first law of thermodynamics for Kerr-anti-de-Sitter
black holes}, Class. Quantum Grav. {\bf 22}, 1503 (2005).

\bibitem{ash07} A. Ashtekar, T. Pawlowski, and C. van den Broeck, 
{\it Mechanics of higher dimensional black holes 
in asymptotically anti-de Sitter spacetimes}, Class. Quantum Grav. {\bf 24}, 625 (2007).

\bibitem {cai99} R. G. Cai and J, H. Cho, {\it 
Thermodynamic curvature of the BTZ black hole}, 
Phys. Rev. D {\bf 60}, 067502 (1999).


\bibitem{jose00} C. Peca, J. P. S. Lemos, {\it Thermodynamics of toroidal black
holes}, J. Math. Physics 41, 4783 (2000). 

\bibitem{callen} H. B. Callen, {\it Thermodynamics and an introduction to 
thermostatics} (John Wiley \& Sons, Inc., New York, 1985).

\bibitem{arnold} V. I. Arnold, {\it Mathematical methods of classical mechanics}
(Springer Verlag, New York, 1980).

\bibitem{rup08} G. Ruppeiner, {\it 
Thermodynamic curvature and phase transitions in Kerr-Newman black holes},
arXiv:gr-qc/0802.1326 


\bibitem{davies} P. C. W. Davies, {\it Thermodynamics of black holes}, Rep. Prog. Phys.
{\bf 41}, 1313 (1978).

\bibitem{dnp86} M. J. Duff, B. E. W.  Nilson, and C. N. Pope, {\it Kaluza-Klein 
supergravity}, Phys. Rep. {\bf 130}, 1 (1986)

\bibitem{jose99} C. Peca, J. P. S. Lemos, {\it Thermodynamics of
Reissner-Nordstr\"om-anti-de Sitter black holes in the grand canonical
ensemble}, Phys. Rev. D 59, 124007 (1999).

\bibitem{cham99} A. Chamblin, R. Emparan, C. V. Johnson, and R. C. Myers, {\it Charged AdS
black holes and catastrophic holography}, Phys. Rev. D {\bf 60}, 064018 (1999).


\bibitem{abb82} L. F. Abbott and S. Deser, {\it Stability of gravity with a 
cosmological constant}, Nucl. Phys. B {\bf 195}, 76 (1982).

\bibitem{hawhor96} S. W. Hawking and G. T. Horowitz, {\it The gravitational 
Hamiltonian, action, entropy and surface terms},  Class. Quantum Grav. 
{\bf 13}, 1487 (1996).

\bibitem{aman06} J. E. \AA man and N. Pidokrajt, {\it Geometry of higher-dimensional
black hole thermodynamics}, Phys. Rev. D {\bf 73}, 024017 (2006). 

\bibitem{cai07} J. Y. Shen, R. G. Cai, B. Wang, and R. K. Su, {\it  
Thermodynamic geometry and critical behavior of black holes}, 
Int. J. Mod. Phys. A {\bf 22}, 11 (2007).


\bibitem{ruppmail} G. Ruppeiner, private communication.

\bibitem{med08} A. J. M. Medved, {\it A Commentary on Ruppeiner metrics for black holes},
arXiv:gr-qc/0801.3497 

\bibitem{coreanos08} Y. S.  Myung, Y. W. Kim, and Y. J. Park, {\it
Ruppeiner geometry and 2D dilaton gravity in the thermodynamics of black holes}, Phys.
Lett. B {\bf 663}, 342 (2008).




\end{thebibliography}
\end{document}